\documentclass[review]{elsarticle}

\usepackage{hyperref,amsmath,bm}
\pdfoutput=1

\usepackage[usenames,dvipsnames]{xcolor}
\definecolor{nblue}{rgb}{0.3,0.3,1.0}
\definecolor{ngreen}{rgb}{0.2,0.7,0.2}
\definecolor{nred}{rgb}{0.9,0.1,0}
\definecolor{nbrown}{rgb}{0.8,0.4,0.15}
\definecolor{nrose}{rgb}{0.7,0,0.35}
\definecolor{nviol}{rgb}{0.5,0,1.0}
\definecolor{nazur}{rgb}{0,0.35,0.7}

\newcommand{\beq}{\begin{equation}}
\newcommand{\eeq}{\end{equation}}
\def\exp{\mathop{\mbox{exp}}}

\newcommand{\DW}{D-Wave 2X}
\newcommand{\tmop}[1]{\ensuremath{\operatorname{#1}}}

\begin{document}

\begin{frontmatter}

\title{A NASA Perspective on Quantum Computing: Opportunities and Challenges}

\author{Rupak Biswas, Zhang Jiang, Kostya Kechezhi, Sergey Knysh, Salvatore Mandr\`a, Bryan O'Gorman, Alejandro Perdomo-Ortiz, Andre Petukhov, John Realpe-G\'omez, 
Eleanor Rieffel, Davide Venturelli, Fedir Vasko, Zhihui Wang
}
\address{NASA Ames Research Center, Moffett Field, CA 94035}

\cortext[mycorrespondingauthor]{Rupak Biswas}
\ead{first.last@nasa.gov}


\begin{abstract}
In the last couple of decades, the world has seen several stunning instances of quantum algorithms that provably outperform the best classical algorithms. For most problems, however, it is currently unknown whether quantum algorithms can provide an advantage, and if so by how much, or how to design quantum algorithms that realize such advantages. Many of the most challenging computational problems arising in the practical world are tackled today by heuristic algorithms that have not been mathematically proven to outperform other approaches but have been shown to be effective empirically. While quantum heuristic algorithms have been proposed, empirical testing becomes possible only as quantum computation hardware is built. The next few years will be exciting as empirical testing of quantum heuristic algorithms becomes more and more feasible. While large-scale universal quantum computers are likely decades away, special-purpose quantum computational hardware has begun to emerge that will become more powerful over time, as well as some small-scale universal quantum computers.
\end{abstract}
\end{frontmatter}

\section{Introduction}

In the last couple of decades, the world has seen several stunning instances of
quantum algorithms that provably outperform the best classical algorithms.  For
most problems, however, it is currently unknown whether quantum algorithms can
provide an advantage, and if so by how much, or how to design quantum
algorithms that realize such advantages.  Many of the most challenging
computational problems arising in the practical world are tackled today by
heuristic algorithms that have not been mathematically proven to outperform
other approaches but have been shown to be effective empirically. While quantum
heuristic algorithms have been proposed, empirical testing becomes possible
only as quantum computation hardware is built. The next few years will be
exciting as empirical testing of quantum heuristic algorithms becomes more and
more feasible.  While large-scale universal quantum computers are likely
decades away, special-purpose quantum computational hardware has begun to
emerge that will become more powerful over time, as well as some small-scale
universal quantum computers.

Successful NASA missions require solution of many challenging computational
problems. The ambitiousness of such future missions depends on our ability to
solve yet more challenging computational problems to support better and greater
autonomy, space vehicle design, rover coordination, air traffic management,
anomaly detection, large data analysis and data fusion, and advanced mission
planning and logistics.  To support NASA's substantial computational needs,
NASA Ames Research Center has a world-class supercomputing facility with one of
the world's most powerful supercomputers.  In 2012, NASA established its
Quantum Artificial Intelligence Laboratory (QuAIL) at Ames to explore the
potential of quantum computing for computational challenges arising in future
agency missions. The following year, through a collaboration with Google and
USRA, NASA hosted one of the earliest quantum annealer prototypes, a 509-qubit
D-Wave II machine, which last summer was upgraded to a 1097-qubit D-Wave 2X
system. 

Because quantum annealers are the most advanced quantum computational hardware
to date, the main focus for the QuAIL team has been on both theoretical and
empirical investigations of quantum annealing, from deeper understanding of the
computational role of certain quantum effects to empirical analyses of quantum
annealer performance on small problems from the domains of planning and
scheduling, fault diagnosis, and machine learning.  This paper will concentrate
on the team's quantum annealing work, with only brief mention of research
related to capabilities of other near-term quantum computational hardware that
will be able to run quantum heuristic algorithms beyond quantum annealing. 
For information on quantum computing more generally, and other algorithms, both
heuristic and non, see quantum computing texts such as \cite{RPbook}.

The power of quantum computation comes from encoding information in a
non-classical way, in qubits, that enable computations to take advantage of
purely quantum effects, such as quantum tunneling, quantum interference, and
quantum entanglement, that are not available classically. The beauty of quantum
annealers is that users can program them without needing to know about the
underlying quantum mechanical effects.  Knowledge of quantum mechanics aids in
more effective programming, just as an understanding of compilation procedures
can aid classical programming, but it is not necessary for a basic
understanding. 

For this reason, the first three sections consist of an overview of quantum
annealing (Sec.~\ref{sec:qaOverview}), a description of how to program a
quantum annealer (Sec.~\ref{sec:qaProgramming}), and a high-level review of our
exploration of three potential application areas for quantum annealing
(Sec.~\ref{sec:qaApps}).  The quantum effects involved are only lightly
mentioned, so these sections should be easily accessible to computer scientists
without any knowledge of quantum mechanics or quantum computing.
Sec.~\ref{sec:qaPhysics}, which examines the role various physical processes
play in quantum annealing, requires more physics knowledge for a full
understanding, as does Sec.~\ref{sec:qaHardware} that discusses hardware,
though a classically-trained computer scientist without knowledge of quantum
mechanics can get a high-level understanding.  We conclude with a brief section
summarizing the outlook for the future.

\section{Quantum annealing}
\label{sec:qaOverview}

Quantum annealing \cite{Farhi-98,Smelyanskiy12} is a metaheuristic optimization
algorithm that makes use of quantum effects such as quantum tunneling and
interference.  It is one of the most accessible quantum algorithms to people
versed in classical computing because of its close ties to classical
optimization algorithms such as simulated annealing and because the most basic
aspects of the algorithm can be captured by a classical cost function and
parameter setting. Quantum annealers are special-purpose quantum computational
devices that can run only the quantum annealing metaheuristic.  For readers not
familiar with quantum annealing in physics, we refer to Sec.~\ref{sec:qaGeneral} for a
general introduction.

Quantum annealers are designed to minimize Quadratic Unconstrained Binary
Optimization (QUBO) problems; i.e., the cost function is of the form
\begin{equation}
C(\mathbf x) = \sum_{i} a_i x_i + \sum_{i < j} b_{i, j} x_ix_ j,
\label{eq:cost_function}
\end{equation}
where $\left\{a_i, b_{i, j} \right\}$ are real coefficients and 
$\mathbf x \in \{0, 1\}^n$
is a vector of binary-valued variables.  An application problem must be mapped
to a QUBO before it can be solved on a quantum annealer.  For application
problems with constraints, the cost function is supplemented with penalty terms
that penalize bit strings that do not correspond to valid solutions. 

The simplicity of the QUBO formalism belies its expressivity.  There exist many
techniques for mapping more complicated problems to QUBO: \begin{itemize} \item
A wide class of optimization problems of practical interest can be expressed in
terms of cost functions that are polynomials over finite sets of binary
variables. Any such function can be re-expressed, through degree-reduction
techniques using ancilla variables, as quadratic functions over binary
variables. We describe such degree-reduction technique in our section on the
CNF mapping of planning problems to QUBO.  \item Cost functions involving
non-binary, but finite-valued, variables can be rewritten in terms of binary
variables alone, and optimization problems with constraints can often be
written entirely in terms of cost functions over binary variables through the
introduction of slack variables.  \end{itemize} For these reasons, the QUBO
setting is more general than it may seem.  We give examples of QUBO mappings
for different applications domains in later sections.

Current quantum annealers such as the \DW\ are fabricated using superconducting
materials and operated at tens of milli-Kelvin temperatures.  The processors
make use of superconducting flux qubits \cite{Harris10} that are superconductor
loops sandwiched with Josephson junctions, engineered so that when an external
flux is applied, a persistent current appears in the loop. The computational
basis of the qubit is the clockwise and counter-clockwise 
flow of the currents, corresponding to values of +1 and -1, respectively, 
of the spin variable $s_j$ for qubit $j$.

An Ising Hamiltonian
\begin{equation}
H_1=\sum_{j} h_j s_j + \sum_{i,j} J_{i,j} s_is_j
\label{eq:Hamiltonian}
\end{equation}
can be programmed on the D-Wave system by setting the values of the flux biases
$h_j$ on each qubit $s_j$ and couplings $J_{i,j}$ between qubits.  A mapping
$s_j=2x_j-1$ relates an Ising Hamiltonian to a QUBO form.  Because only select
couplers are implemented in the hardware, only certain quadratic terms can be
directly implemented. Embedding, using multiple qubits to represent a single
binary variable, is necessary to implement arbitrary QUBOs, a topic we will
return to when we discuss programming quantum annealers in more depth. 

\begin{figure*}[t]
\begin{center}
\includegraphics[height=0.6\textwidth]{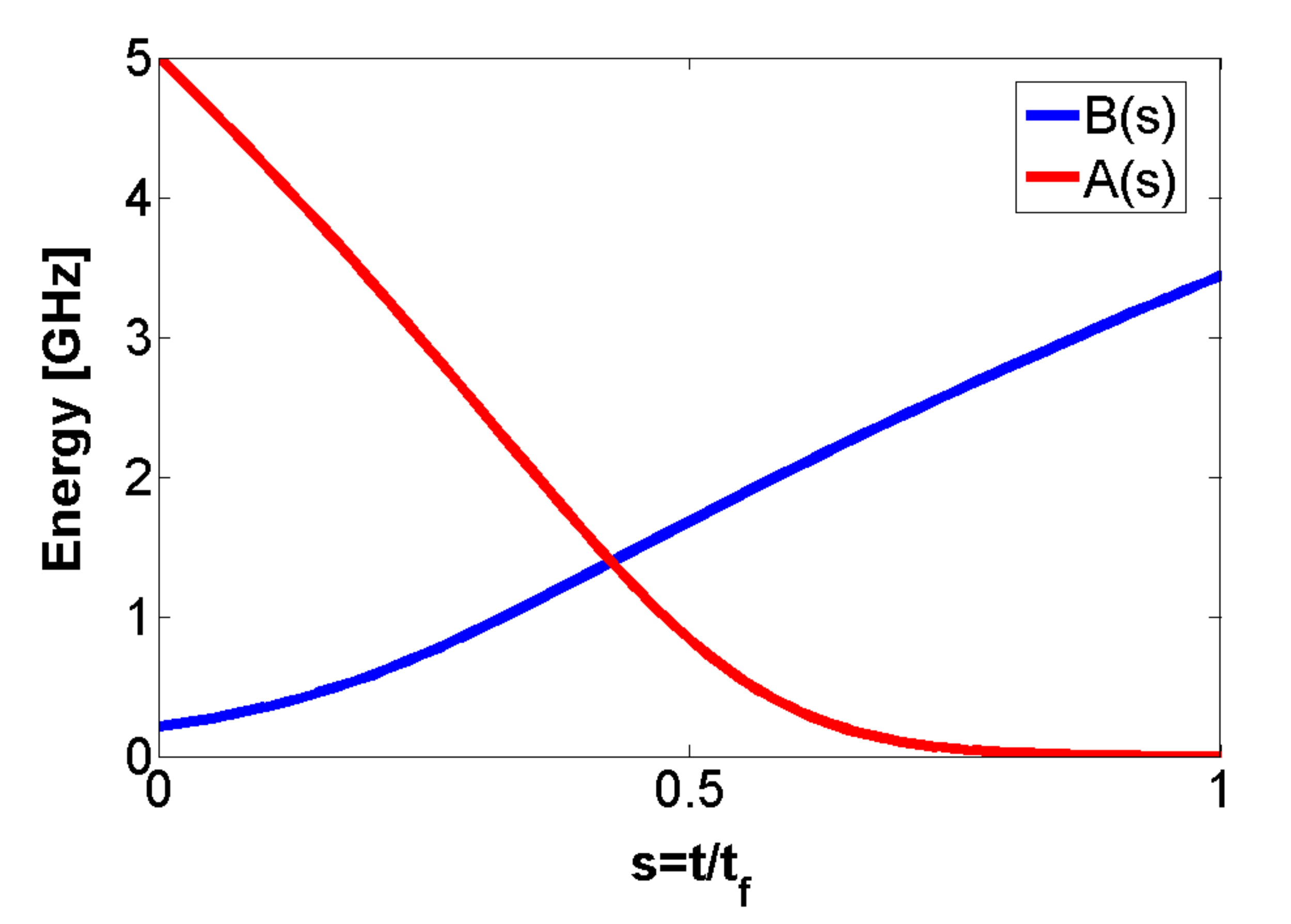}
\caption{
Typical annealing profile $A(s)$ and $B(s)$.} \label{fig:annealSch}
\end{center}
\end{figure*}

Quantum annealing is carried out by evolving the system under the
time-dependent Hamiltonian 
 \begin{equation}
 H(t)=A(s) H_0+B(s) H_1
 \end{equation}
where $H_1$ is the problem Hamiltonian in QUBO form and $H_0$ is
the initial Hamiltonian, which in current annealers is fixed and cannot be set
by the programmer. Generally, the Hamiltonian $H_0$ is chosen to have a simple
energy landscape so that an unsophisticated relaxation process will efficiently
put the system in low energy states.  During the anneal, $H_0$ is gradually
changed until it becomes $H_1$.  The intuition is that if the system starts in
low energy states and the change is smooth enough, the system will end up in
low energy states of the final Hamiltonian, just as a top spinning on a tray
will continue to spin when the tray is moved as long as the change in position
is smooth enough.  The functions $A(s)$ and $B(s)$ are generally chosen in a
way that $H_0$ dominates at $s=0$ and $H_1$ dominates at $s=1$ (see
Fig.~\ref{fig:annealSch}).  Current annealers provide a range of total anneal
times $t_f$, where $s = t/t_f$, enabling traversals at different speeds.  On
the \DW\ housed at NASA, the annealing time can be chosen in a range from
$5~\mu$s to $2$~ms. Future annealers may allow programmers to choose $A(s)$ and
$B(s)$, but they are currently fixed in the \DW.

When viewed as an algorithm for exploring the landscape defined by the cost
function to find a global minimum, quantum annealing resembles a commonly used
classical algorithm for optimization: simulated annealing. While in simulated
annealing thermal fluctuation provides the mobility over energy barriers
between local minima, quantum annealing has an additional source of mobility:
quantum fluctuations that facilitate tunneling through the barriers.  Such
quantum fluctuations are realized through $H_0$ which serves as a driver
Hamiltonian responsible for quantum fluctuations because it does not commute
with the target Hamiltonian $H_1$.  As the anneal continues, the driver term is
reduced, slowly turning off the fluctuations, as the problem Hamiltonian's
strength increases. 

Quantum annealing should not be confused with adiabatic quantum computation
which is known to support universal quantum computing.  The problem Hamiltonian
in quantum annealing typically is a classical Hamiltonian.  While adiabatic
quantum computation also interpolates between an initial and final Hamiltonian,
the final Hamiltonian can be highly non-classical with no analogous classical
cost function, thus enabling much more general sorts of quantum computations. 

\section{Programming a quantum annealer}
\label{sec:qaProgramming}

This section discusses the two main steps in programming a quantum annealer:
{\it mapping\/} the problems to QUBO; and {\it embedding\/}, which takes these
hardware-independent QUBOs to other QUBOs that match the specific quantum
annealing hardware that will be used.

\subsection{Mapping}
For a cost function not natively in QUBO form, the typical procedure to map the
problem into QUBO is  to properly choose binary variables, formulate
constraints, and embed the violation of constraints as energy penalties.  We
illustrate this process with an example from Ref.~\cite{Rieffel15a}. 

{\bf Example:} In a graph coloring problem, the task is to determine whether
each vertex of a graph $G(V,E)$ can be colored from a set $\cal C$ so that no
two vertices connected by an edge have the same color.  The goal is to
formulate a cost function such that the minimum is 0.  One way to choose the
binary variable is to use $x_{v,c}=0$ or $1$ to express whether vertex $v$ is
assigned color $c$.  The ensuing constraints would be: (1) Each vertex needs to
be assigned exactly one color that can be expressed in binary form as $(\sum_c
x_{v,c}-1)^2$. (2) Connected vertices cannot share the same color; otherwise,
the energy penalty is raised,  $\sum_c \sum_{v,v'\in E} x_{v,c}x_{v',c}$.  The
cost function expressed in QUBO is then $H=\sum_v (\sum_c x_{v,c}-1)^2+\sum_c
\sum_{v,v'\in E} x_{v,c}x_{v',c}$.  When no requirement is violated, the cost
function has value 0, which is the ground state of $H$.

In this example, the cost function $H$ is naturally quadratic.  More generally,
the cost functions of many optimization problems can be expressed as
higher-degree polynomials of the binary variables (PUBOs).  Degree-reduction
techniques can then be applied to recast a PUBO as QUBO, usually at the price
of adding ancilla variables \cite{Babbush13}.
	
\subsection{Embedding}
Because the physical hardware has limited connectivity, there usually does not
exist a direct one-to-one mapping between the QUBO binary variables and the
physical qubits so that each binary term in the QUBO corresponds to a pair of
connected qubits.  To obtain the needed connectivity in the embeddable QUBO, an
additional step is required.  Unlike the mapping step, the embedding step is
hardware dependent.  A cluster of qubits $\{y_{i,k}\}$ connected to each other
in the hardware graph will represent a single variable $x_i$.  For any term
$x_ix_j$ in the mapped QUBO, there is a connection in the embeddable QUBO
between one of the qubits in the cluster for $x_i$ and one qubit in the cluster
for $x_j$.  Minor embedding is the process of determining a cluster for each
binary variable in the problem QUBO \cite{Choi2011}.  The problem of finding
the optimal minor embedding is itself NP-complete, but fortunately it is not
necessary to find the optimal embedding. 
In general, for planar architectures,
there are straightforward, fast algorithms to embed an $N$-variable problem in
hardware consisting of no more than $N^2$ physical qubits
\cite{Choi2011,Klymko14,zaribafiyan2016}.  In the near term, while the hardware
is so qubit constrained, heuristic algorithms \cite{cai2014practical} are used to try to
minimize resources and maximize the size of the problems embeddable on the
machine.

To encourage the qubits in the cluster to all take the same value by the end of
the anneal so that the value of the variable they represent is unambiguous, the
embeddable QUBO also includes constraint terms $J_F y_{i,p}y_{i,q}$ for any
pair $p, q$ of qubits in the cluster that are connected to each other, where
$J_F$ is the strength of the coupling. This is to ensure that in the most
energy-favorable configuration, all qubits in the cluster take the same value.
The Hamiltonian obtained from the embeddable QUBO shares the same ground state
energy as the Hamiltonian from the mapped QUBO, but conforms to the hardware
architecture.  The higher energy spectrum may be considerably altered, so
different embeddings can significantly affect performance.

The optimal strength of $J_F$ is a subject of extensive research
\cite{Rieffel15a,OGorman2015,Venturelli15a}.  One might think it should be as
high as possible to force the qubits to all take the same value at the end, but
in practice there is a sweet spot.  Coupling strengths that are too high
degrade performance. Intuitively, a high coupling strength makes it harder to
change the value of a variable in the cluster once they take on a value that is
not, ultimately, optimal, though the actual quantum dynamics are more
complicated than this simple explanation.

\begin{figure*}[t]
\begin{center}
\includegraphics[width=0.6\textwidth]{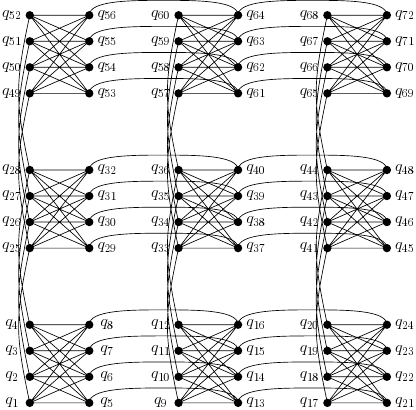}
\caption{Nine unit cells in a Chimera graph.}
\label{fig:Chimera}
\end{center}
\end{figure*}

\begin{figure*}[t]
\begin{center}
\includegraphics[height=0.2\textwidth]{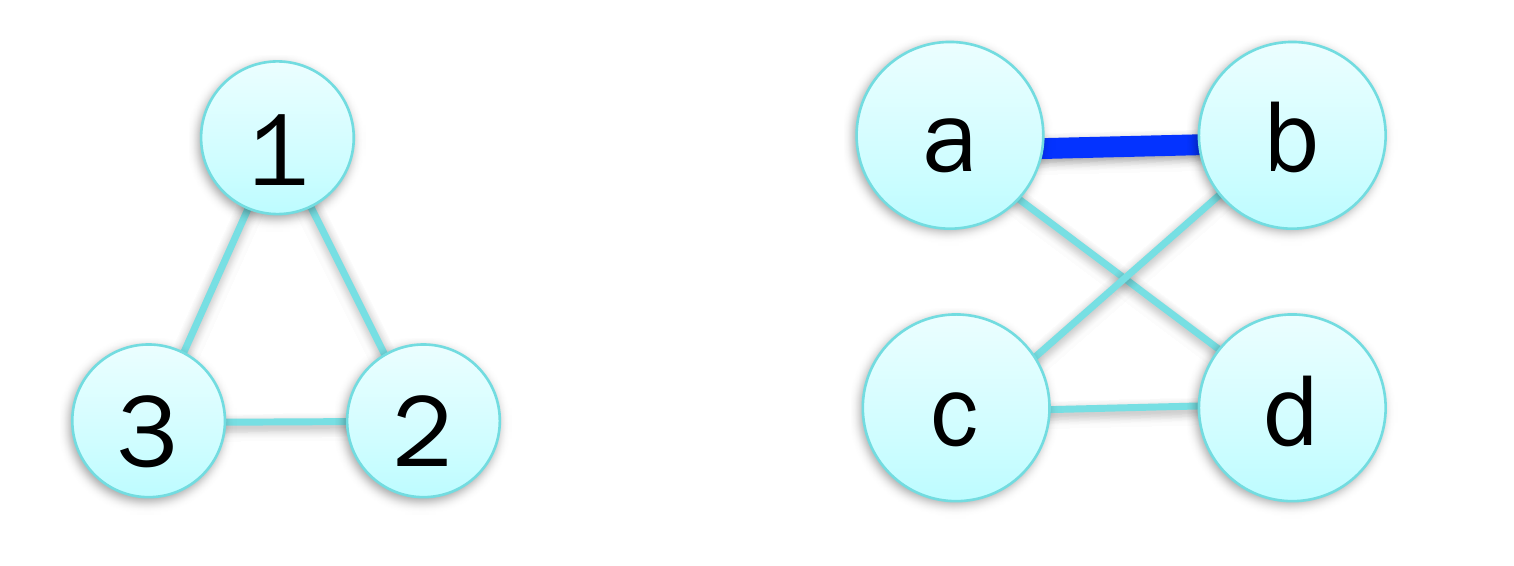}
\caption{
Schematics of embedding the Hamiltonian $H=J_{1,2}s_1 s_2+J_{1,3}s_1 s_3+ J_{2,3} s_2 s_3$ on a graph. 
Left: Triangle graph to be embedded.  Right: Graph after embedding on a
bi-partite graph of size 4. The variable $s_1$ is represented by two physical
qubits $s_a$ and $s_b$ with a strong ferro-magnetic coupling $J_F <0$.  The
Hamiltonian after embedding is 
$H_{\text {embed}}=J_F s_a s_b + J_{1,2} s_a s_d + J_{1,3} s_b s_c + J_{2,3} s_c s_d$.
}
\label{fig:embedding}
\end{center}
\end{figure*}

The layout of the qubits and couplers of a D-Wave quantum annealer is a
$n\times n$ lattice of unit cells called a Chimera graph.  Each unit cell is
composed of a bi-partite graph of 8 qubits.  A schematic diagram  of the graph
formed by 9 cells is shown in Fig.~\ref{fig:Chimera}.  The current D-Wave
machine at NASA has $12\times 12$ such units and a total of 1152 qubits, of
which 1097 are working. Each qubit is coupled to at most 6 other qubits, 4
within its own unit cell and 2 to qubits in its neighboring cells.  To embed a
generic QUBO of $N$ variables, $N^2$ qubits and couplers are needed in the
worst case so that each binary variable can be represented by $N$ physical
qubits and effectively couple to all other binary variables.  As an
illustration, Fig.~\ref{fig:embedding} shows an example of embedding a triangle
onto a bi-partite graph.

When an Ising problem is programmed to the chip, errors due to noise or
manufacturing miscalibration associated with the bias fields ($h$'s) and
couplers ($J$'s) would affect the annealing performance. Simple offset errors
can be corrected through software, but more complicated errors are harder to
mitigate. One strategy is to repeat the annealing with a gauge-transformed
Hamiltonian in which the states used to represent $0$ and $1$ are swapped. The
qubits are encoded into $s'_j=g_j s_j$ where $g_j=\pm 1$, and the biases and
couplers are accordingly set as $h'_j=g_jh_j$ and $J'_{i,j}=g_ig_jJ_{i,j}$.
The resulting Hamiltonian $H'=\sum_j h'_j s'_j + \sum_{i,j} J'_{i,j} s'_i
s'_j$, which is equal to the original Hamiltonian, is sent to the annealer and
the solution obtained is then decoded using $s_j=g_j s'_j$.  One set of
parameters $\{g_j\}$ is called a gauge. In the absence of errors, the annealing
results for $H$ and $H'$ should be the same while the actual performance could
be gauge-dependent.  Success probabilities averaged over a set of gauges are
typically used. Various error suppression and correction strategies exist, both
fully quantum \cite{jordan2006}, a mix of quantum and classical~\cite{vinci2015}, 
and a more recent quantum approach \cite{Jiang2015}.  Once
the problem is programmed, the annealing is repeated multiple times (typically
thousands to millions), and each time the final state measured in the
computational basis is recorded.

\section{Applications}
\label{sec:qaApps}

In this section, we give a high-level overview of our in-depth studies of three
potential applications areas: planning and scheduling, fault diagnosis, and
machine learning. Further technical details can be found in the publications
referenced in each section.

\subsection{Quantum annealing for planning and scheduling}

Automated planning and scheduling has many applications, from logistics, air
traffic control, and industrial automation to conventional 
military missions, resource allocation, and assistance in disaster recovery.  
Many of the challenges in autonomous
operations include significant planning and multi-agent coordination tasks in
which operational teams must generate courses of action prior to the event and
adjust those plans as new information becomes available or unexpected events
occur. 

Many planning and scheduling problems are very challenging to solve; as the
number of events to plan or schedule grows, the number of possible solutions grows
exponentially.  These problems are often NP-hard or harder, and are currently
tackled by classical heuristic algorithms.  The emergence of quantum annealing
hardware allows the exploration of quantum heuristic approaches to these
problems~\cite{Smelyanskiy12}, with the objectives to search for significant
improvements over existing techniques in the efficiency with which good plans
can be found, or in finding better plans that satisfy more constraints, and/or
in greater diversity in the plans found.  

Given the severe limitation in quantum memory of current quantum annealers, in
order to benchmark the machines, it is imperative to find prescriptions to
identify small problems that exhibit signature of hardness.  Currently, the
most common approach to designing benchmark planning problems is to extract
solvable problems from real-world applications. This approach has the benefit
of tuning algorithms toward the applications from which the benchmark problems
are obtained. A complementary approach is to design parametrized families that
capture aspects of practical planning problems and can be shown to be
intrinsically hard. Such families support focused examination of these aspects,
small problems that can be meaningfully considered to be hard, and scaling
analyses with respect to size.  Families of small but hard problems are
critical for present research into quantum annealing because the current
quantum annealers can handle only small problems. Families we have designed for
the purpose of assessing the performance of quantum annealers have proved
useful in distinguishing the strengths and weaknesses of state-of-the-art
planners~\cite{Rieffel14}.

\subsubsection{QUBO formulation of general planning problems}

\begin{figure*}[t]
\begin{center}
\includegraphics[width=0.8\textwidth]{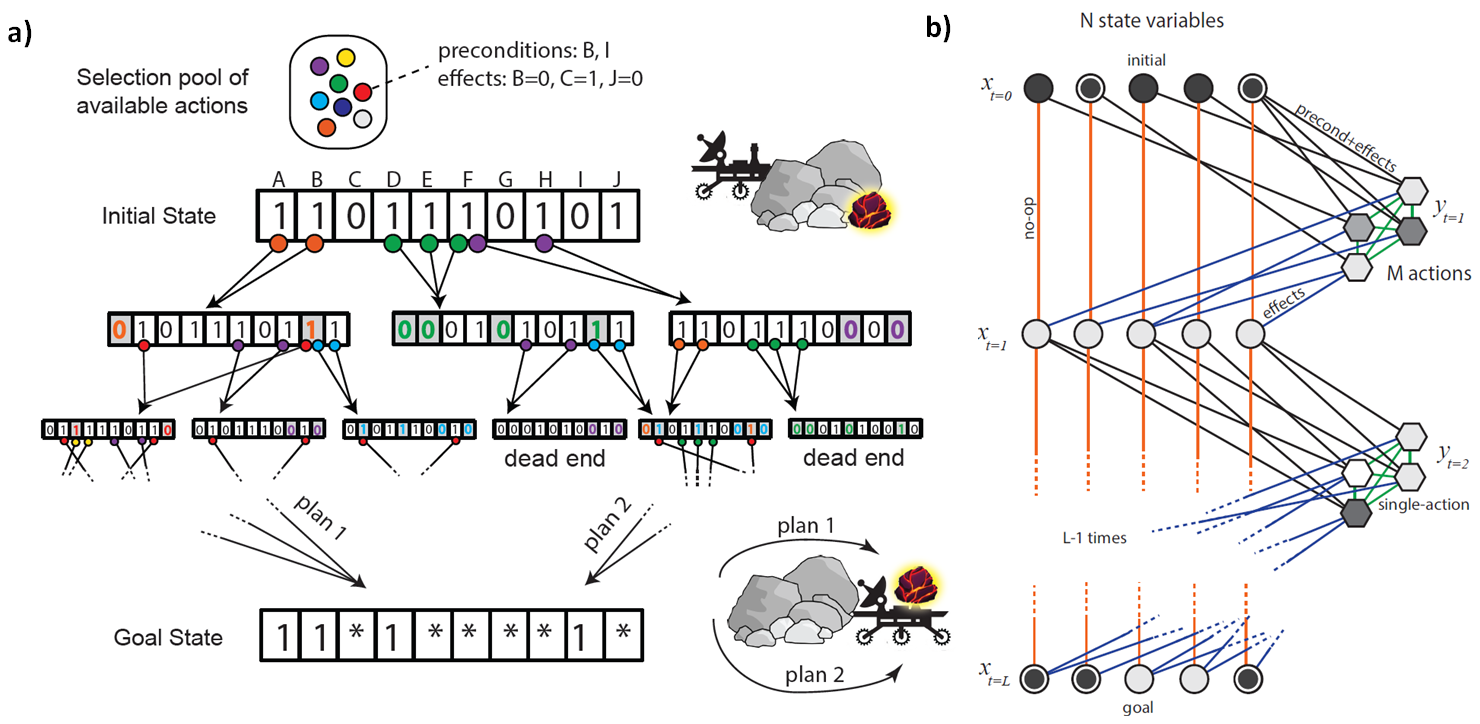}
\caption{(a) Pictorial view of a planning problem. The initial state (e.g.,
Rover behind the rocks without sample) is specified by assigning True (1) or
False (0) to state variables (named A-J in this simplified example). The
planning software navigates a tree, where a path represents a sequence (with
possible repetitions) of actions selected from a pool (colors). Each action has
preconditions on the state variables (e.g., moves can be done around the rocks
but not through them) that need to be satisfied for the actions to be executed
(the circles under the state variables in the search tree need to be True) and
has an effect on the state (colored variables in shaded regions of the new
state have changed values). A valid search plan (multiple valid plans are
possible) will reach the goal state (e.g., Rover in front of the rocks with a
sample collected).  (b) Direct time-indexed QUBO structure for a planning
problem with only positive preconditions and goals.  Each node represents a
state variable (left) or an action (right) at any given time $t$. Time flows
from top to bottom, and variables $y_i^{(t)}$ for the actions at time $t$ are
shown between the state variables $x_i^{(t-1)}$ for one time step and the state
variables $x_i^{(t)}$ for the next time step. The node grayscale intensity
represents the magnitude of local field (bias) $h_i$ applied to a given qubit
$i$, and the double contour in a node indicates a negative bias.}
\label{fig:planningOverview}
\end{center}
\end{figure*}

Classical planning problems are expressed in terms of binary {\it state
variables\/} and {\it actions}.  Examples of state 
variables in the domain of autonomous rover navigation are
``Rover $R$ is in location $X$" and ``Rover $R$ has a soil sample
from location $X$," which may be True or False. Actions consist of two lists, a
set of {\it preconditions} and a set of {\it effects} (see
Fig.~\ref{fig:planningOverview}).  The effects of an action consists of a
subset of state variables with the values they take on if the action is carried
out. For example, the action ``Rover $R$ moves from location $X$ to location
$Y$" has one precondition, ``Rover $R$ is in location $X$ = True" and has two
effects ``Rover $R$ is in location $X$ = False" and ``Rover $R$ is in location
$Y$ = True."

A specific planning problem specifies an {\it initial state\/}, with values
specified for all state variables, and a {\it goal\/}, specified values for one
or more state variables. As for preconditions, goals are conventionally
positive, so the specified value for the goal variables is True. Generally, the
goal specifies values for only a small subset of the state variables. A plan is
a sequence of actions. A valid plan, or a solution to the planning problem, is
a sequence of actions $A_1, ..., A_L$ such that the state at time step
$t_{i-1}$ meets the preconditions for action $A_i$, the effects of action $A_i$
are reflected in the state at time step $t_i$, and the state at the end has all
of the goal variables set to True.

Ref.~\cite{Rieffel15a} discusses a general QUBO formulation of planning
problems (see Fig.~\ref{fig:planningOverview}(b)). If the original planning
problem has $N$ state variables and we are looking for a plan of length $L$,
then the QUBO problem will have $N(L + 1)$ binary variables $x_i^{(t)}$, where
$t\in\{0,\dots,L\}$ is the time index, and $i$ is the index of the state
variable in the original planning problem. In addition, if the original
planning problem has $M$ possible actions, we will have $LM$ additional binary
variables $y_j^{(t)}$ which indicate whether the $j$th action is carried out at
time step $t$ or not. A QUBO can then be defined in terms of these variables,
with terms  capturing the goal, precondition, effect, single-action, and no-op
(no variable change without an action) constraints:
\begin{equation}
 H = H'_{\tmop{goal}} + H_{\tmop{no} - \tmop{op}}
 + H'_{\tmop{precond}} + H_{\tmop{effects}} + H_{\tmop{single-action}}.
\end{equation}
Ref.~\cite{Rieffel15a} describes a somewhat more general cost function that 
supports multiple actions per time step.

\subsubsection{Advanced scheduling applications}

Scheduling was recognized early on as one the most promising near-term targets
for quantum annealing due to its efficient quadratic time-indexed Mixed-Integer
Linear Programming formulation. Furthermore, there is a rich literature of
complex pre-processing and hybrid classical techniques. Using this direct
quadratic formulation of scheduling instead of the most general planning
formulation leads to very significant performance advantages in runs of the
D-Wave machines~\cite{Rieffel15a}. 

Scheduling formalizes problems dealing with the optimal allocation of resources
(machines, people) to tasks (jobs) over time, under various constraints and
figures of merit. In one direct QUBO formulation, a bit is associated to the
execution of a given job in a given machine (out of $M$ possible) at a given
time (discretized in $T$ slots), allowing for very efficient mappings on
current quantum annealers supporting two-body Ising-type interactions, using
$NMT$ qubits, where $N$ is the number of jobs. While objective functions of the
{\it priority maximization\/} type are easily implementable as linear penalty
functions requiring only local fields on the corresponding logical bits,
objectives requiring {\it makespan minimization\/} require a more involved
encoding with either $T$ ancilla clock variables highly connected to the qubits
relative to the jobs scheduled last, or by complementing the quantum solver
with guidance from classical methods, such as binary
search~\cite{Venturelli15a}.

Many planning and scheduling problems are of such scale and complexity that
they are by necessity solved in pieces, and so quantum hardware can be
naturally integrated into the solution of such problems.  Hybrid solvers
employing quantum annealing together with classical methods are particularly
suited to scheduling applications, because the state-of-the-art approaches for
specific scheduling problems are typically combining different approaches in a
modular way, and decompositions can be employed to get around programming
bottlenecks such as high connectivity, precision requirements, continuous
constraints, or to employ quantum annealing as a heuristic module of a complete
solver~\cite{Tran16a,Tran16b}.  As a heuristic module of a complete solver,
quantum annealing enables more directed search of the solution space.  Building
a complete solver out of a probabilistic quantum subroutine requires
non-trivial classical co-processing, but recent work has shown that it can be
done successfully.  In particular, partial solutions returned by a quantum
solver can be used to derive bounds on the optimum value of the function to
be optimized, 
and therefore focus on the most promising or neglect the least promising parts of the solution space.

Recent work on the application of quantum annealing to scheduling includes
programming and benchmarking quantum annealers on small problems from the
domains of graph coloring~\cite{Rieffel15a}, job shop
scheduling~\cite{Venturelli15a}, Mars lander activity
scheduling~\cite{Tran16a}, air traffic runway landing~\cite{Tran16b}, and
alternative resource scheduling~\cite{Tran16b}.  The question of speedup with
respect to purely classical methods are inconclusive due to the small size of
the problems implementable on current quantum annealers and the inefficiency
of embedding techniques~\cite{Rieffel15a}.  This body of work has identified
precision and connectivity requirements that suggest future generations of
annealers may be able to solve currently intractable scheduling problems within
a decade. 

Planned technological advances in quantum annealing architectures will also
make possible tighter integration of quantum and classical components in the
hybrid approaches discussed above, both through more programmable devices that
allow for greater flexibility as subroutines and through application-specific
devices that maximize the effectiveness of particular algorithms.  
In future, we expect quantum hardware to be integrated into larger systems much as
graphical processing units are today~\cite{dasari2016programmable}.

\subsection{Fault detection and diagnostics of graph-based systems}

Another application domain we have studied with quantum annealing devices is
the diagnostics of electrical power-distribution systems (EPS); a collaboration
between QuAIL and the Discovery and System Health (DaSH) technical area at NASA
Ames. Diagnosing the minimal number of faults capable of explaining a set of
given observations, e.g., from sensor readouts, is a hard combinatorial
optimization problem usually addressed with artificial intelligence techniques.
In~\cite{PerdomoOrtiz_EPJST2015}, we presented the first application of the
Combinatorial Problem $\rightarrow$ QUBO Mapping $\rightarrow$ Direct Embedding
process where we were able to embed instances with sizes comparable to those
found in real-world problems.  We demonstrated problem instances with over 100
electrical components (including circuit breakers and sensors) running on a
quantum annealing device with 509 quantum bits. In comparison, the number of
components in the electrical circuits used for diagnostics competitions from
NASA's Advanced Diagnostics and Prognostics Testbed (ADAPT) ranges between 40
and 100~\cite{kurtoglu09first}. 

\subsubsection{QUBO formulation}
\label{subsubsec:FDmapping2qubo}

\begin{figure*}
\centering
\includegraphics[width=0.95\textwidth]{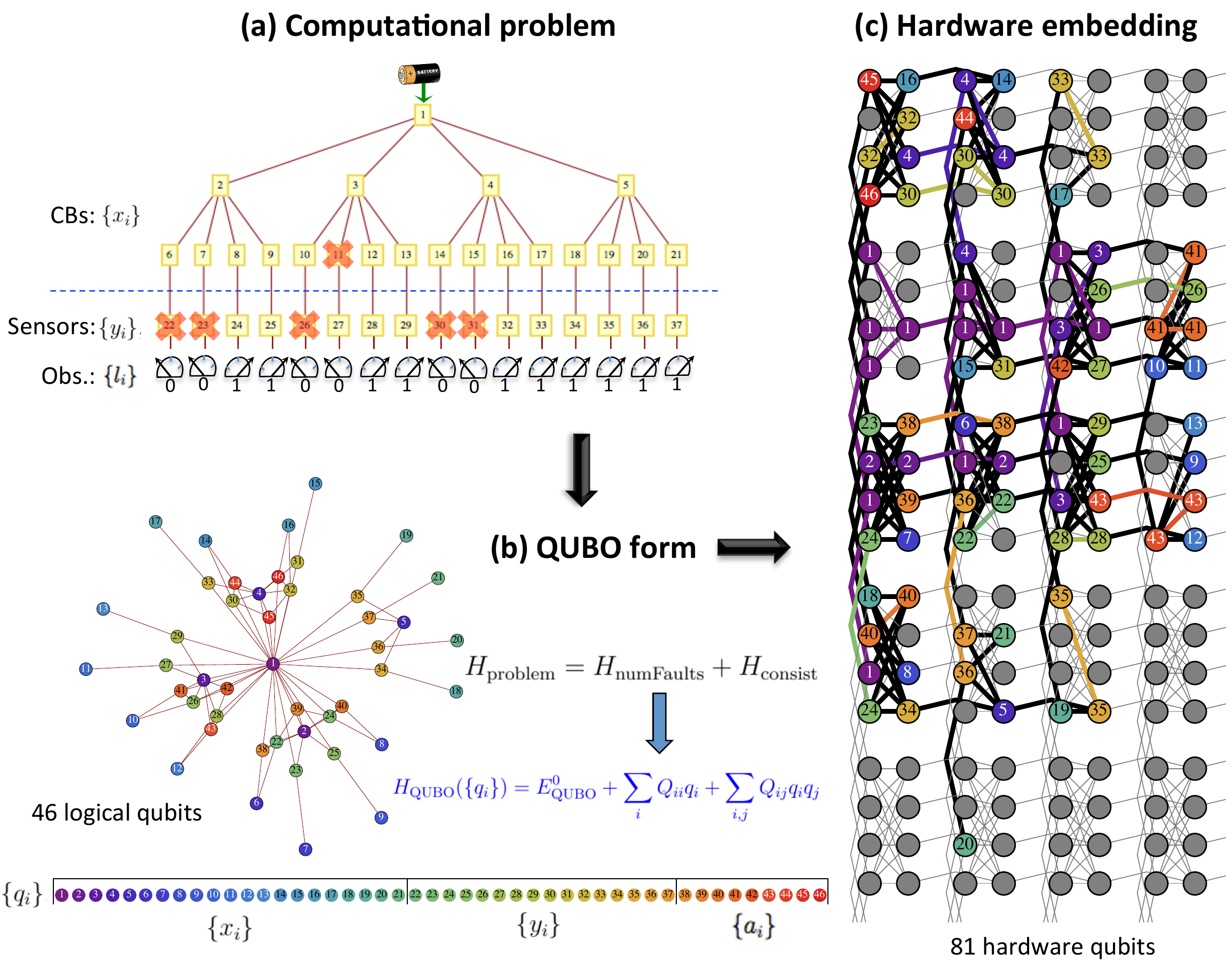}
\caption{General scheme of an experimental setup for the diagnosis of multiple
faults with a quantum annealer. (a)~A possible realization of the diagnosis of
multiple faults in an EPS network with one power source, 21 CBs and 16 sensors
or ammeters. The orange crosses indicate faulty electrical components
($x_i=0)$. In this particular instance of 6 faults, a plausible explanation
of the readouts places one of the faults on a CB and the remaining 5 on the
ammeters. However, this is only one of the $2^6$ six-fault explanations that
are equally likely in this case. (b)~QUBO form of the problem where coupling
between two logical qubits is represented as edges. (c)~The subsequent
embedding into the Chimera graph usually requires more variables since some
logical qubits are represented by several physical qubits (depicted here as
nodes in the graph) due to the sparse connectivity of the hardware graph. In
this problem, 81 physical qubits are needed to implement the QUBO with 46
logical variables.} \label{fig:diagnosis}
\end{figure*}

As shown in Fig.~\ref{fig:diagnosis}(a), there are two types of components.
The first are circuit breakers (CB), which in their healthy mode allow the flow of current,
and are illustrated as the nodes of the quaternary tree. We denote them by the
set of binary variables $\{x_i\}$, with $x_i=1$ ($x_i=0$) corresponding to CB
$i$ in a healthy (faulty) state. The other component type is the sensor or
ammeter, which is not only another electrical component that could potentially
malfunction, but also forms part of the observations from which one is asked to
perform the diagnosis of the electrical network. Therefore, for each ammeter,
we have an observation parameter and a status variable indicating its healthy
or faulty state.  The observations (or readouts) are part of the problem
definition and provided as input parameters. We denote this set of binary
parameters $\{l_i\}$, with $l_i=1$ ($l_i=0$) if the $i$-th ammeter is showing a
High (Low) readout. Similar to the $\{x_i\}$ variables for the CBs, the
uncertainty in the ammeter readouts is introduced by assigning to them a set of
binary variables, $\{y_i\}$, with $y_i=1$ ($y_i=0$) corresponding to ammeter
$i$ in a healthy (faulty) state.

The goal is to find the minimum number of faults in the electrical components,
either in the CBs and/or the ammeters, consistent with the circuit layout and
the readouts. We solve this as a minimization problem over the pseudo-Boolean
function $H_{\text{problem}}(\{x_i\},\{y_i\};\{l_i\})$, whose construction is
explained below. After $H_{\text{problem}}$ is transformed into its QUBO form,
we can subsequently use the quantum annealer to find the assignment for each of
the $\{x_i\}$ and $\{y_i\}$. 

The construction of the pseudo-Boolean function contains two contributions:
\begin{equation}
\label{eq:Hproblem}
H_{\text{problem}} = H_{\text{numFaults}} + H_{\text{consist}}.
\end{equation}
$H_{\text{consist}}$ is constructed such that it is 0 whenever the
prediction from the assignment of all the $\{x_i\}$ and $\{y_i\}$ is consistent
with the readouts $\{l_i\}$ from the ammeters, and greater than 0 when the
readouts and the prediction, given the $\{x_i\}$ and $\{y_i\}$ assignments, do
not match. Consider the set $P_i$ as the set of CB indices in the path from the
root node (CB 1) where power is input, all the way to the CB connected to the
$i$-th ammeter. Thus, for the network in Fig.~\ref{fig:diagnosis}(a), $P_1 =
\{1,2,6\}$, $P_2 = \{1,2,7\}$, $\cdots$, and $P_{16} = \{1,5,21\}$. If we
denote the number of paths as $n_{\text{paths}}$ (equals the number of ammeters
in this network), one can construct $H_{\text{consist}}$ as:
\begin{equation}\label{eq:Hconsist}
H_{\text{consist}} = \lambda_{\text{path}} \sum_{i=1}^{n_{\text{paths}}}  y_i g_i, \quad   f_{i}(\{x_j\}_{j \in P_i}) = \prod_{j\in P_i} x_j,
\end{equation}
with $g_i = l_i + f_{i}-2f_i l_i$, a binary function with $g_i=0$ when the
prediction $f_i$, based only on the CB statuses in the path $P_i$, is
consistent with the readouts $l_i$, and $g_i=1$ when the prediction and the
readout are in disagreement. In other words, $g_i = \textsc{xor}(f_i,l_i)$.

$H_{\text{numFaults}}$ is proportional to the number of faults (whenever $x_i = 0$ or $y_i = 0$)
in the electrical network:
\begin{equation}
\label{eq:Hfaults}
H_{\text{numFaults}} = \lambda^{\text{CB}}_{\text{faults}} \sum_{i=1}^{n_{\text{CB}}} (1-x_i)+ \lambda^{\text{sensor}}_{\text{faults}} \sum_{i=1}^{n_{\text{sensor}}} (1-y_i),
\end{equation}
and when combined with $H_{\text{consist}}$, as written in
Eq.~(\ref{eq:Hproblem}), defines the problem energy function to be minimized by
favoring the minimal set of faulty components that are simultaneously
consistent with the observations in the outermost sensors. A thorough
discussion on setting the values of all the penalties is provided
in~\cite{PerdomoOrtiz_EPJST2015}.

Notice the pseudo-Boolean $H_{\text{consist}}$ is a high-degree polynomial, and
for this particular network, the order of the polynomial is related to the
depth of the tree. We can reduce the degree of the polynomial to a quadratic
expression, $H_{\text{QUBO}}$, with the overhead of adding more binary
variables, while conserving the global minimum of the original function,
$H(\{x_i\},\{y_i\};\{l_i\})$. Further details on the techniques used for this
reduction are provided in~\cite{PerdomoOrtiz_EPJST2015,Babbush2012}. 

Assuming it requires $n_A$ ancilla variables $\{a_i\}$ to reduce the
high-degree polynomial to the quadratic expression, we can relabel the CB,
sensor, and ancilla variables, $\{x_i\}$, $\{y_i\}$, and $\{a_i\}$,
respectively, into a new set of binary variables $\{q_i\}$ for $i = 1,2,
\cdots, n_{l}$, with $n_{l} = n_{\text{CB}}+ n_{\text{sensor}}+n_{A}$ as the
total number of logical qubits. The final quadratic cost function to be
minimized can then be written as
\begin{equation}\label{eq:Hqubo}
\begin{split}
H_{\text{QUBO}}(\{q_i\}) &= E^0_{\text{QUBO}} + \sum_{i,j}  Q_{i,j} q_i q_j \\&= E^0_{\text{QUBO}} + \mathbf{q}^T \cdot \mathbf{Q} \cdot \mathbf{q}.
\end{split}
\end{equation}
As shown in Fig.~\ref{fig:diagnosis}, this expression can be represented as a
graph with the number of vertices equal to the number of logical qubits $n_{l}$
corresponding to the set of variables $\{q_i\}$. In this representation,
$Q_{i,i}$ can be treated as the weights on the vertices, while $Q_{i,j}$ are the
weights for the edges representing the couplings between variables $i$ and $j$
(see Fig.~\ref{fig:diagnosis}). Notice that since $q^2_i = q_i$, the expression
$\mathbf{q}^T \cdot \mathbf{Q} \cdot \mathbf{q}$ contains both linear terms
$Q_{i,i}$, and quadratic terms, $Q_{i,j}$, when $i \neq j$. $E^0_{\text{QUBO}}$
corresponds to the constant independent term.

Although the problems studied in~\cite{PerdomoOrtiz_EPJST2015} are simpler than
typical real-world instances, we believe that they still capture some
non-trivial features, such as the inclusion of uncertainty in the sensor
readouts. Of course, aiming to embed all the details from realistic scenarios
will require significantly more qubits and also depend on the specific
network/problem to be solved.

As another realization of the fault detection application, the QuAIL team is
examining combinational digital circuits~\cite{Feldman2007}, a more realistic
scenario used to benchmark codes devoted to solving diagnostics related
problems~\cite{kurtoglu09first}. Preliminary results look
very promising and harder than any other benchmarks reported in the literature
and used to address the question of quantum speedup in quantum annealers.

\subsection{Sampling and machine learning applications}

Sampling from high-dimensional probability distributions is at the core of a
wide spectrum of computational techniques with important applications across
science, engineering, and society. Examples include deep learning,
probabilistic programming, and other machine learning and artificial
intelligence applications. 

Much of the record-breaking performance of classical machine learning
algorithms regularly reported in the literature pertains to task-specific
supervised learning algorithms~\cite{Bengio-Book}.  Unsupervised learning
algorithms are more human-like, and in principle more general and powerful, but
their development has been lagging due to the intractability of traditional
sampling techniques such as Markov Chain Monte Carlo (MCMC).  Indeed, as
leading researchers in the field have pointed out~\cite{Bengio-Book}, future
success of unsupervised learning algorithms requires breakthroughs in efficient
sampling algorithms.  Quantum annealing holds the potential to sample more
efficiently and from more complex probabilistic models, which would
significantly advance the field of unsupervised learning.

\subsubsection{A different class of problems for quantum annealing}

A computationally hard problem, key for some relevant machine learning tasks,
is the estimation of averages over probabilistic models defined in terms of a
Boltzmann distribution \beq\label{e:PB}
P_B(\mathbf{s})=\frac{1}{Z}\exp\left({\sum_{i,j}  W_{ij} s_i s_j + \sum_{i} b_i
s_i}\right), \eeq where $Z$ is the normalization constant or partition
function, $\mathbf{s} =\{s_1,\dots ,s_N\}$ denotes a configuration of binary
variables, and $W_{ij}$ and $b_i$ are the parameters specifying the probability
distribution. 

Sampling from generic probabilistic models, such as $P_B(\mathbf{s})$ in
Eq.~\eqref{e:PB}, is hard~\cite{Frigessi-Biometrika-1997} in general.  For this
reason, algorithms relying heavily on sampling are expected to remain
intractable no matter how large and powerful classical computing resources
become. Even though quantum annealers were designed for challenging
combinatorial optimization problems, it has been recently recognized as a
potential candidate to speed up computations that rely on sampling by
exploiting quantum effects, such as quantum tunneling
~\cite{Benedetti-2016,Amin-arXiv-2015}. 

\subsubsection{Quantum-assisted learning of Boltzmann machines}

Indeed, some research groups have recently explored the use of quantum
annealing hardware for the learning of Boltzmann machines and deep neural
networks (see \cite{Benedetti-2016,Adachi-arXiv-2015} and references therein).
The standard approach to the learning of Boltzmann machines relies on the
computation of certain averages that can be estimated by standard sampling
techniques, such as MCMC.  Another possibility is to rely on a physical
process, like quantum annealing, that naturally generates samples from a
Boltzmann distribution.  In contrast to their use for optimization, when
applying quantum annealing hardware to the learning of Boltzmann machines, the
control parameters (instead of the qubits' states) are the relevant variables
of the problem. The objective is to find the optimal control parameters that
best represent the empirical distribution of a given dataset. 

These ideas are framed within a hybrid quantum-classical computing paradigm.
Given a classical machine learning infrastructure, the idea is to replace the
software module that generate samples, e.g., via MCMC, with a quantum annealing
process. This quantum sampling module could be similarly employed in other
domains where sampling is useful. Thus, demonstrating quantum speedup for
sampling would have broad implications. 

In recent work~\cite{Benedetti-2016}, the QuAIL team has demonstrated how to
properly use a quantum annealer by overcoming critical challenges such as the
instances-dependent temperature estimation. In fact, while the probability
distribution $P_B(\mathbf{s})$ in Eq.~\eqref{e:PB} is specified by parameters
$W_{ij}$ and $b_i$, the control parameters of a quantum annealer are instead
$J_{ij}=T_{\rm eff}\, W_{ij}$ and $h_i = T_{\rm eff}\, b_i$. According to
quantum dynamical arguments~\cite{Amin-arXiv-2015}, $T_{\rm eff}$ is an {\it
instance-dependent\/} effective temperature, different from the physical
temperature of the device. Unveiling this unknown temperature is key to
effectively using a quantum annealer for Boltzmann sampling.  By introducing a
simple effective temperature estimation algorithm \cite{Benedetti-2016}, it was
possible to successfully use the D-Wave 2X system for the learning of a special
class of restricted Boltzmann machines that can serve as a building block for
deep learning architectures. Experiments run using a synthetic dataset showed
that the quantum-assisted algorithm outperformed in terms of quality (i.e., the
value of the likelihood reached) the standard classical algorithm named CD-1
and approached the performance of CD-100, which takes about 100 times more
computational effort than CD-1 (See~\cite{Benedetti-2016} for details).
Complementary work that appeared roughly simultaneously showed that quantum
annealing can be used for supervised learning in classification
tasks~\cite{Adachi-arXiv-2015}. 

These results are encouraging, but there remain numerous challenges before the
full potential of quantum annealing hardware for sampling problems can be
harnessed.  While each future generation will no doubt be an improvement,
hardware advances alone will not suffice.  The QuAIL team is therefore
developing algorithmic strategies to address these other problems, with
promising initial results. For example, we recently 
experimentally demonstrated~\cite{Benedetti-2016b} the feasibility of a fully unsupervised machine
learning application by successfully training our quantum annealer, using up to
940 qubits, to generate, reconstruct, and classify images that closely resemble
(low resolution) handwritten digits, among other synthetic datasets. We showed
a Turing test (see Fig. 4 in \cite{Benedetti-2016b}) to challenge people to
distinguish between handwritten digits and digits generated by the quantum
device; most people we informally showed this Turing test either failed or
found it difficult. To reach this milestone, we implemented densely connected
hardware-embedded models that are more robust to noise and more efficient to
learn with state-of-the-art quantum annealers. 

The ultimate question that drives this endeavor is whether there is quantum
speedup in sampling applications. Current experience with the use of quantum
annealers for combinatorial optimization suggest the answer is not
straightforward. This work is part of the emerging field of quantum machine
learning~\cite{Schuld-QML-2015}, an essentially unexplored territory where
quantum annealing might have a large impact in the near term. 

\subsection{Best practice programming and compilation techniques}

These explorations have spurred QuAIL to design advanced techniques to guide
programming and improve performance.  Software calibration methods devised by
the team are described in~\cite{PerdomoOrtiz_SciRep2016}.
In~\cite{Rieffel15a}, we compare different mappings and in~\cite{Perdomo15a},
we present advanced techniques to intelligently select gauges based on small
numbers of trial runs that often improve performance by an order of magnitude.
Compilation strategies for quantum annealers, including guidelines for
optimally setting the  strength of $J_F$ are discussed in
\cite{Rieffel15a,OGorman2015,Venturelli15a}.  
Furthermore, we have identified certain common structures in the QUBO representations of many applications because different constraints often have similar forms~\cite{Rieffel15a}.

\section{Physics of quantum annealing}
\label{sec:qaPhysics}
This section discusses results clarifying the role of various processes in
quantum annealing that suggest where to look for potential quantum speedup and
where such an advantage would be unlikely.  So far, we have been informal about
what we mean by quantum speedup.  However, knowing the different types of quantum
speedup is helpful in assessing results related to the computational power of
quantum annealing. It is also necessary to improve our understanding of
potential classes of problems for which such a quantum device can excel.

\subsection{Background on quantum annealing}
\label{sec:qaGeneral}

The target of quantum annealing is to optimize a function of QUBO form, as in
Eq.~(\ref{eq:cost_function}).  The cost function has a physical realization in
a system comprising quantum bits (qubits) where each binary variable is encoded as a
qubit. The coefficients ($a_i$) translate into bias fields applied on the
qubits and ($b_{i,j}$) is represented as the coupling strength between two
qubits.  The cost function thus corresponds to a {\it Hamiltonian}, $H_1$, as
in Eq.~(\ref{eq:Hamiltonian}), which describes the energy of the system.  The
Hamiltonian bears strong similarity with the cost function.  However, while in
the classical cost function the binary variables can take value either 0 or
1, in a Hamiltonian the qubit is allowed to be (and in a physical quantum
system, can be) in a superposition of these two states $\alpha |0\rangle +
\beta |1\rangle$.  The optimization problem translates into finding the ground
state of the Hamiltonian, i.e., the eigenstate of the lowest eigenvalue of
$H_1$.  In order to do so, quantum annealing introduces quantum fluctuation in
the system, represented as a non-commuting term in the Hamiltonian, $H_0$.  A
typical $H_0$ easy to prepare physically is $H_0=\sum_j \sigma^x_j$ where each
$\sigma^x_j$ swaps states $0$ and $1$ on the $j$-th qubit.  The weight of $H_0$
with respect to $H_1$ is the strength of the fluctuation.   The initial state of
the system is one with all possible classical configurations that are equally likely.
The system starts with a strong quantum fluctuation that gradually
quenches. The quantum fluctuation provided by $H_0$ allows the dynamics to
explore a larger region of the search space and gradually concentrate (with large
probability) at the global minimum. 
At the beginning of the search, the initial state is very far from the global minimum but a large fluctuation allows the system a better chance to accept a state that is energetically higher; thus allowing a more extensive search of the solution space.
As the annealing progresses, the fluctuation is tuned down and the system spends more and more time around the global minimum, eventually staying there once the fluctuation disappears.
This process resembles simulated annealing where the
quantum fluctuation replaces the thermal fluctuations. 

Another perspective of the same process is to view the 
total Hamiltonian as slow moving and time dependent.
If the Hamiltonian is varying slowly
enough, the system will follow its instantaneous eigen-state 
(this is known as the adiabatic theorem).   Since the initial state is actually
the ground state of $H_0$, a slow tuning would eventually result in the ground
state of the problem Hamiltonian, $H_1$.  A key question is: how slow is slow
enough? During the evolution when there is another energy level close to the
ground state and if the change of Hamiltonian is not slow enough, there is a risk
the system would jump to the higher level and never return, and the algorithm
would fail.  The closer the two energy levels are, the slower the Hamiltonian must
vary in order to mitigate this risk.  The spectral gap (the minimal distance
between the two energy levels) plays a crucial role in quantum annealing. 
 
Ref.~\cite{ronnow2014defining} defines four classes of quantum speedups:
\begin{itemize}
\item {\it Provable quantum speedup:\/} It is rigorously proven that no
classical algorithm can scale better than a given quantum algorithm. 
\item {\it Strong quantum speedup:\/} The quantum heuristic is faster than
any known classical algorithm. This type of speedup has been established for
dozens of special-purpose algorithms, with Shor's polynomial-time 
algorithm for factorization being the most prominent. The {\it best\/}
classical algorithm  
may be continually evolving, as is the case for most areas in which classical 
heuristics prevail; the ICAPS (International Conference on Automated Planning
and Scheduling) planning competition and the SAT competition
generally see new algorithms every year.
\item {\it Potential quantum speedup:\/} The quantum speedup is in comparison to
a specific classical algorithm or a set of classical
algorithms. 
\item {\it Limited quantum speedup:\/} There is a quantum speedup
only if the quantum heuristic is compared to the closest classical counterpart.
\end{itemize}
A finer-grained classification, which takes into account the
type of classical algorithm used in the comparison, has been proposed
in \cite{mandra2016strengths}. 

To better understand where quantum annealing may confer an advantage, it is
important to appreciate its major sources of error. The algorithm may fail to
find a solution due to escape from the ground state via 
either non-adiabatic transitions or decoherence processes. Yet another
possibility is that the ground state does not correspond to the optimal
solution due to control noise. In the following, we review some of the
recent developments in assessing the impact of these error mechanisms.

\subsection{Quantum annealing bottlenecks} Some insight into the relative
performance of quantum annealing can be gained by studying random optimization
problems using the tools of the the statistical mechanics. Absent noise,
non-adiabatic transitions can be prevented only if the annealing proceeds
slowly across points where the gap $\Delta E$ that separates the instantaneous
ground state from excited states becomes small (taking at least time $t \propto
\hbar / \Delta E$). The most widely discussed bottleneck, where the gap reaches
a local minimum, is the quantum phase transition. Some of the computationally
hardest problems exhibit a discontinuous (first order) phase transition, where
the gap is exponentially small. In a common scenario, the ground state
wavefunction abruptly changes from being a superposition of a large number of
spin configurations to being nearly localized near a global minimum. If the
transverse field is lowered too fast, the algorithm performs no better than a
random guess.

Continuous (second order) phase transitions scale better, although strong
fluctuations of disorder (randomness of the parameters of the problem) can
still make the gap scale as a stretched exponential (exponential in some
fractional power of problem size). This still leaves a large swath of problems
--- most amenable to quantum annealing --- where the disorder is irrelevant
at the critical point (phase transition) so that the gap there
is only polynomially small.  Recent work \cite{knysh2016} addresses this
practically relevant scenario and finds that after the phase transition
bottleneck, the algorithm encounters further bottlenecks with gaps that scale as
a stretched exponential.

As annealing progresses, the number of spin configurations with significant
amplitudes decreases until the wavefunction is completely localized. This is
roughly equivalent to having a partial assignment of variables: An increasing
fraction of binary variables have converged to a definite value, while the
remaining variables are in a superposition state. At times, a state with a
different assignment of already fixed variables becomes more energetically
favorable, and a large number of variable have to flip simultaneously in a
multi-qubit tunneling, which is the source of "hard" bottlenecks described
above.  This process is analogous to "backtracking" in classical search
algorithms.

The major finding is that the number of tunneling bottlenecks is proportional
to the logarithm of problem size. In practice, as the problem size increases,
the time complexity of quantum annealing will exhibit a crossover from
polynomial scaling (when the phase transition bottleneck is dominant) to
exponential (when the expected number of "hard" bottlenecks exceeds one).  This
size threshold is related to the "density" of spin glass bottlenecks.  Similar
concept can be introduced for other heurstic search algorithms, such as
simulated annealing. The bottleneck density can thus be used as a metric of
performance indicating problem sizes above which the time complexity increases
exponentially.

Interestingly, the minimum requirement for the annealing time is to avoid
non-adiabatic transition at the phase transition (polynomial scaling).  As it
turns out, for fully coherent annealing, having one long annealing cycle versus
choosing the best out of repeated short cycles results in identical time-to-solution 
(as long as annealing time exceeds the aforementioned minimum).
Shorter annealing times minimize the effects of decoherence and have been
favored in most experimental studies on the D-Wave hardware.

Coupling to the environment affects these results in multiple ways.
First, it changes the universality class of the phase transition, worsening
scaling of the minimum annealing cycle \cite{takada2016}. Second, it suppresses
multi-qubit tunneling since in addition to flipping qubits, corresponding
environmental degrees of freedom have to adjust. If quantum-mechanical
tunneling is strongly suppressed, equilibrium may be reached via thermal
excitation due to finite temperature. In this regime, performance would
paradoxically improve with increasing temperature as the system becomes more
classical.

\subsection{Multi-qubit co-tunneling} Multi-qubit quantum co-tunneling is
expected to be a key microscopic mechanism responsible for quantum speedup in
quantum annealing. In the following, we consider limited speedup; i.e., speedup
compared to simulated annealing.  Realistic hardware is subject to intrinsic
noise that affects the quantum dynamics of the system, and therefore needs to
be considered when evaluating the efficiency of quantum annealing hardware. The
effect of hardware noise is twofold: (1)~Coupling to noise allows inelastic
processes, prohibited by energy conservation in the closed system. Inelastic
relaxation provides an efficient route to a local minimum within a
convex region of the potential energy landscape. (2)~Dephasing noise leads to
loss of coherence between the states on different sides of the barrier,
resulting in an incoherent tunneling regime, and, in the strong coupling
regime, causes renormalization of the tunneling rate. 

In the case of the flux qubits of the D-Wave system, the typical
decoherence time (a measure of how long quantum features of a single qubit can
be maintained, specifically the characteristic decay time of the off-diagonal
elements of the qubit’s density matrix) is of the order of nanoseconds to tens
of nanoseconds, which is shorter than the minimum run time of the annealing
schedule, 5 microseconds. Nevertheless, D-wave annealers demonstrate signatures
of quantum many-body dynamics, particularly incoherent multi-qubit quantum
tunneling and evidence of 8-qubit tunneling has been reported~\cite{
SmelyanskiyGoogle14}. In the course of quantum annealing, the dynamics of the
device is limited to low-energy multi-qubit superposition states, which are
more robust against the effects of noise and decoherence than single qubit
states. In this regime, single qubit excitations caused by noise local to each
qubit are strongly suppressed by the strong qubit-qubit coupling energy. At the
same time, slow fluctuations of local magnetic flux result in a time-dependent
spectrum of the multi-qubit low-energy states, which introduces decoherence of
the multi-qubit dynamics.

In the vicinity of the algorithm's bottlenecks, quantum annealing hardware
realizes incoherent tunneling~\cite{SmelyanskiyGoogle14}.  Different tunneling
regimes are determined by comparing the quantum tunneling rate near the
computational bottleneck to the characteristic dephasing rate. In a common
regime, the tunneling rate near the bottleneck is exponentially small, while
the dephasing rate is at least of order one. In this regime, quantum tunneling
can be only incoherent in nature~\cite{Leggett87}: an analog of the decay of a
metastable state into a continuous spectrum encountered in nuclear physics and
chemistry, as opposed to a coherent superposition of states on two sides of a
potential barrier. The incoherent regime is characterized by a quadratic
slowdown of quantum tunneling.  Nevertheless, there exist classes of problems
where limited polynomial speedup is possible in this regime, particularly in
cases where the shape of the potential barrier favors quantum tunneling over
classical over-the-barrier escape, such as when barriers are tall and
thin~\cite{GoogleTunneling2015}.

An alternative~\cite{Kechedzhi16}, operational also in the case of thick
barriers where the usual intuition would favor classical 
escape, is the class of problems characterized by exponential degeneracy of the
metastable state separated by a barrier from the global minima. The latter is
typical for NP-hard problems; a common feature of classical mean-field spin
glass models~\cite{Cavagna97} is a polynomial number of the global minima separated
by large potential barriers from an exponential number of metastable states. In
such a landscape, simulated annealing slows down exponentially due to an
additional entropic barrier associated with escaping the exponentially
degenerate set of metastable states. In contrast, in the course of quantum
annealing, the transverse field splits the degeneracy of the classical problem
and thereby avoids the additional entropic barrier.

To better understand multi-qubit tunneling processes, we developed an
instantonic calculus for analytical treatment of the thermally-assisted
tunneling decay rate of metastable states in fully-connected quantum spin
models~\cite{jorg2010energy, Semerjian-wkb}. The tunneling decay problem can be
mapped onto the Kramers escape problem of a classical random dynamical field.
This dynamical field is simulated efficiently by path integral Quantum Monte
Carlo (QMC). We show analytically that the exponential scaling with the number
of spins of the thermally-assisted quantum tunneling rate and the escape rate
of the QMC process are identical~\cite{jiang2016scaling}. This analytical
result complements prior numerical work~\cite{isakov2015understanding} and
provides an explanatory model.  This effect is due to the existence of a
dominant instantonic tunneling path.  We solve exactly the nonlinear dynamical
mean-field theory equations for a single-site magnetization vector that
describe this instanton trajectory.  We also derive scaling relations for the
``spiky'' barrier shape when the spin tunneling and QMC rates scale
polynomially with the number of spins while a classical over-the-barrier
activation rate scales exponentially.

\subsection{Role of noise} Intrinsic noise cannot be eliminated from real
quantum devices: manufacturing imperfections, as well as thermal fluctuations,
induce quantum dephasing and decoherence (see Section~\ref{sec:qaHardware}).
Noise can sometimes be helpful (thermal fluctuations are responsible for the
thermally-assisted annealing effects discussed earlier), but can cause quantum
devices to work far from their ideal state, limiting the actual performance and
hiding any potential quantum speedup.

In addition, control noise can change the target Hamiltonian $H_1$ with the
consequence that the target solution is no longer in the ground subspace of
$H_1$. In this case, even a perfect quantum device, subject only to control
noise, would find a ``false'' ground state, which could be far from any target
solution.  The maximum noise that can be added to $H_1$ before the target
solutions do not belong to the ground subspace of $H_1$ is called {\it
resilience\/}~\cite{Venturelli15,katzgraber2015seeking}. In general, resilience
can be increased by properly rescaling the energy of $H_1$.  Real quantum
devices, however, have a limited range of energies so the resilience cannot be
completely neglected. Recent work shows that a low resilience could hide a
quantum speedup~\cite{Venturelli15}.

\section{Quantum annealing hardware} \label{sec:qaHardware} To date, the most
significant progress in quantum annealing hardware is based on the engineering
of quantum superconducting circuits with macroscopic collective variables
(e.g., electric charge and magnetic flux) exhibiting quantum coherence. Here we
review basic design and operational principles of such circuits, focusing on
different types of superconducting qubits, inter-qubit coupling, and
decoherence processes caused by various sources of the environmental noise. 

\subsection{Quantization of electric circuits with Josephson junctions}

\begin{figure}
\begin{center}
\includegraphics[scale=0.5]{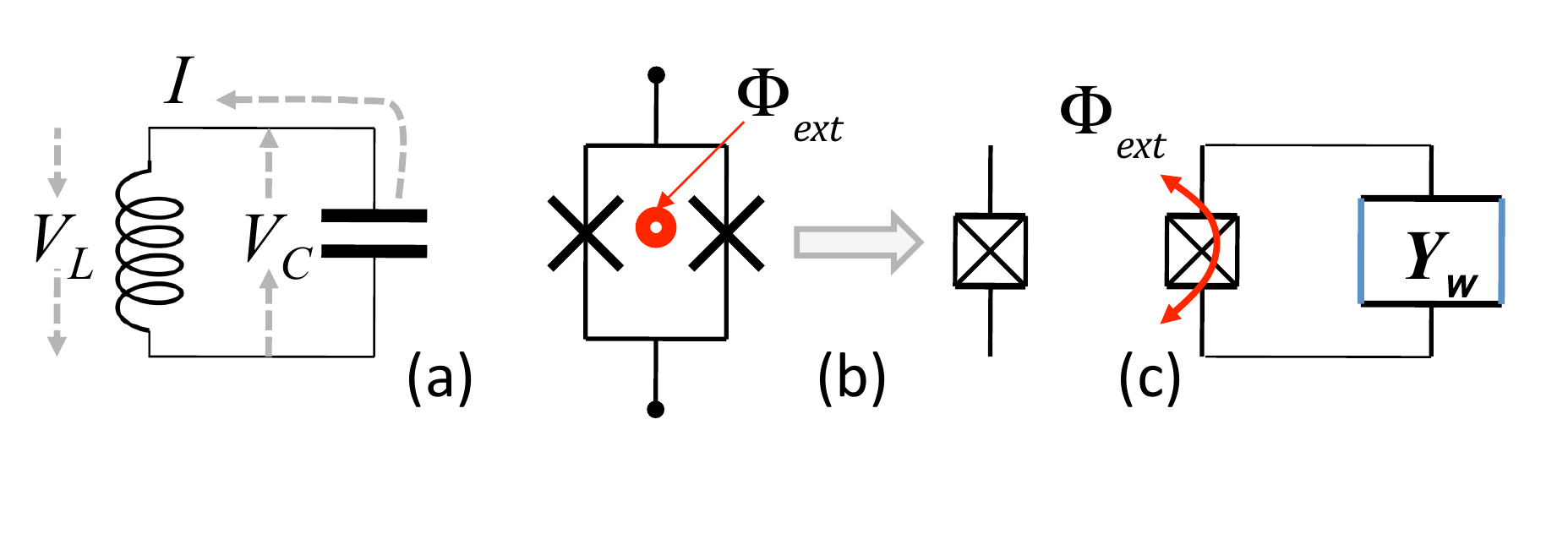}
-\end{center}\addvspace{-0.5 cm}
\caption{(a)~Lumped element model for LC oscillator with current $I$ and voltages
$V_C=-V_J$. (b)~Tunable SQUID loop biased by external flux $\Phi_{ext}$.
(c)~Effective circuit of a qubit.}
\label{Fig:Circuits}
\end{figure}

Let us briefly describe quantization of zero-resistance superconducting
circuits, which is based on the lumped element
method~\cite{devoret:1997,Devoret:2004,Schoelkopf:2008cs}.  We can represent a
circuit using two alternative sets of variables: current and voltage ($I(t)$
and $V(t)$) or charge and flux ($Q(t)$ and $\Phi(t)$), connected with each
other via the relations $I=dQ/dt$ and $V=d\Phi/dt$.  Let us start with the
simplest circuit such as an LC oscillator (see Fig.~\ref{Fig:Circuits}(a)),
whose dynamics is governed by the Kirchhoff's laws $I_{L} =I_{C}\equiv I$ and
$V_{L} +V_{C} =0$. Using $V_{L}=LdI/dt$ and $V_{C}=Q_{C}/C$, one obtains the
equation of motion
$ \ddot{I}+\omega_{LC}^2I = 0$, where $\omega _{LC}  = 1/{\sqrt{LC}}$ 
is the characteristic frequency for classical current (and voltage) oscillations.
The magnetic flux  $\Phi$ and charge
$Q$ are governed by similar equations, e.g., $\ddot{\Phi}+\omega_{LC}^2\Phi=0$.
Using variables $(Q,\Phi)$ one can express the equations of motion in the
Hamiltonian form, $\dot\Phi=\partial H/\partial Q$ and $\dot Q=-\partial
H/\partial\Phi$, where the classical Hamiltonian function is
$H=Q^2/2C+\Phi^2/2L$.  Following the standard quantization procedure, we
replace classical variables with corresponding operators, introduce the
commutator $\left[{\hat\Phi ,\hat Q}\right] =i\hbar$, and arrive at the
Hamiltonian of a quantum harmonic oscillator, $\hat H =\hat Q^2 /2C+\hat\Phi^2
/2L$, describing the quantized electromagnetic modes of a macroscopic LC
circuit with equidistant energy levels, $E_n  = \hbar \omega _{LC} (n + 1/2)$
with $n = 1,2\ldots$. Clearly, this energy spectrum is not suitable for an
implementation of a {\em two-level} qubit.

In order to separate two well-defined levels that can be used as logical states
$|0\rangle$ and  $|1\rangle$, one should employ a non-harmonic circuit with
almost negligible coupling of the qubit levels and the rest of the spectrum. A
natural solution is to introduce a Josephson junction as a nonlinear and
non-dissipative element of the circuit. Josephson junctions are formed by two
superconductors weakly connected through a high barrier. Within the lumped
element approach, they are described by the current-voltage characteristics
$I_J =I_0\sin (2\pi\Phi /\Phi_0)$ where $I_0$ is a critical current and $\Phi_0
=\pi\hbar /e$ is the flux quantum. Analysis of different realizations of a
qubit is based on the Kirchhoff's laws and on the description of the junction's
contributions in terms of $I_{J}$ and $V_{J}$ (or $\Phi$). 

A tunable qubit is realized if one replaces a single Josephson junction by a
SQUID loop formed by two parallel junctions biased by an external flux,
$\Phi_{ext}$ (see Fig.~\ref{Fig:Circuits}(b)).  The current passing through the
SQUID is given by 
$
I_J =I_0\cos (2\pi\Phi_{ext}/\Phi_0)\sin (2\pi\Phi /\Phi_0 )
$
\cite{Harris:2010},
which can be thought of as an effective junction with tunable critical current
$I_{\text{eff}}=I_0\cos (2\pi\Phi_{ext}/\Phi_0)$ controlled by $\Phi_{ext}$. A typical
tunable qubit can be represented as an effective  junction shunted by a linear
circuit with admittance $Y_\omega$ (see Fig.~\ref{Fig:Circuits}(c)). Below we
consider two basic types of  such qubits shunted by either LC oscillator (flux
qubit) or a capacitor (charge qubit). 

\subsection{Hamiltonians of flux and charge qubits}

\begin{figure}
\begin{center}
\includegraphics[scale=0.5]{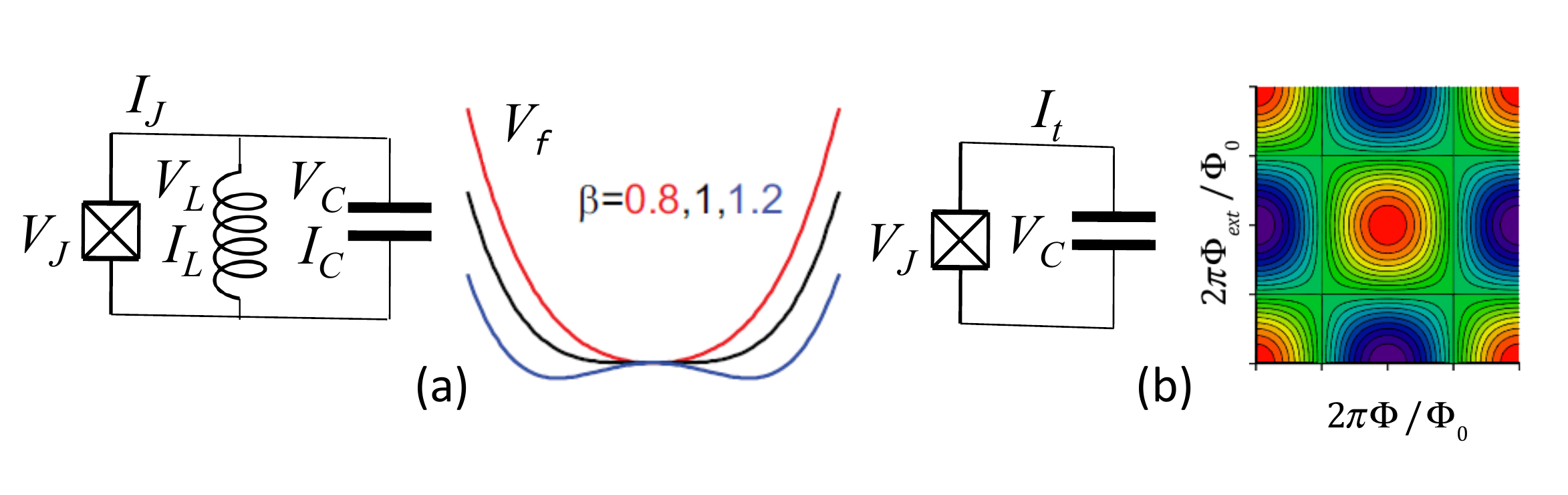}
\end{center}\addvspace{-0.5 cm}
\caption{Effective circuits and potential energies vs.\ flux for: (a)~Flux qubit (tunable
Josephson junction shunted by LC oscillator); and (b)~Junction shunted by capacitor only (charge
qubit).}
\label{Fig:Qubits}
\end{figure}

Effective Josephson junctions, inductance and capacitance, connected in
parallel, form a {\em flux qubit} (see Fig.~\ref{Fig:Qubits}(a)). The circuit
is governed by the Kirchhoff's laws for currents $I_{J,L,C}$ and voltages
$V_{J,L,C}$: $I_J =I_C +I_L$, $V_J +V_C =0$, and $V_J +V_L =0$.
Using these relations, we obtain the equation of motion for  a flux $\Phi$
threading through the device as:
$C{{\ddot \Phi}} + {{\Phi}}/{L} +I_{\text{eff}} \sin\left( \frac{{2\pi \Phi }}{{\Phi _0 }}\right) = 0$,   which leads to the following
Hamiltonian of a flux qubit
\begin{equation}
\label{Ham1}
\hat H  = \frac{\hat Q^2}{2C} + \frac{{(\hat\Phi-\Phi_x)^2 }}{{2L}} - \frac{{\Phi _0 I_{\text{eff}} }}{{2\pi }}\cos \frac{{2\pi \hat\Phi }}{{\Phi _0 }} ~. 
\end{equation}
Here we assumed that the inductance loop $L$ can be biased by  an additional
external flux $\Phi_x$ applied through inductive coupling. The first
(capacitance) term $\hat Q^2 /2C$  in Eq.~(\ref{Ham1}) can be interpreted as a
kinetic energy while the second and third terms describe a potential formed by
inductance and Josephson terms, respectively. 

For further consideration, it is convenient to introduce dimensionless flux
$\hat\phi =2\pi\hat\Phi /\Phi_0 +\pi$ and charge operators $\hat
q=-id/d\hat\phi$. Then the Hamiltonian in Eq.~(\ref{Ham1}) can be expressed as
\begin{equation}
\label{Ham2}
\hat H = 4E_C \hat q^2  + E_L \frac{(\hat\phi -\phi_x)^2}{2} + E_J^{\text{eff}} \cos\hat\phi ,
\end{equation}
and it is different from the LC oscillator by adding the effective energy of
Josephson junction, $E_J^{\text{eff}}=\Phi _0 I_{\text{eff}}/2\pi$. We also
introduce here the capacitance and inductance energies, $E_C=e^2 /2C$ and $E_L
=(\Phi_0 /2\pi)^2 /L$ and $\phi_x=2\pi\Phi_x/\Phi_0+\pi$.  The Hamiltonian in
Eq.~(\ref{Ham2}) corresponds to a particle with kinetic energy proportional to
$E_C$ and potential energy determined by the interplay between $E_L$ and
$E_J^{\text{eff}}$ through the ratio $\beta =E_J^{\text{eff}}/E_L =2\pi
I_{\text{eff}}L/\Phi_0$.  If $\beta < 1$, Eq.~(\ref{Ham2}) describes a
single-well anharmonic oscillator, while for $\beta >1$ the double-well
potential emerges and there are two closely-spaced tunnel-split levels defining
a qubit. The quantum dynamics is determined by tunneling between the wells that
can be controlled by variation of the barrier height $E_J^{\text{eff}}$, and by
tilting the two-well potential via the tilt flux $\phi_x$.  Flux qubits
described by Eq.~(\ref{Ham2}) are implemented in D-Wave quantum
annealers~\cite{Harris:2010}.

A typical {\it charge qubit\/} operates as an open circuit shown in
Fig.~\ref{Fig:Qubits}(b).  To derive the Hamiltonian we must omit the
inductance term in the equations of motion, which results in
\begin{equation}
\label{Ham3}
\hat H = 4E_C \hat q^2 + E_J^{\text{eff}}\cos \phi,
\end{equation}
and contains only the Josephson (periodic) part of the potential energy. The
eigenvalue problem is reduced to the Mathieu equation. Operational regimes of
various qubits described by the generic Hamiltonian in Eq.~(\ref{Ham3})
drastically depend on the ratio  $E_J/E_C$.

\begin{figure}
\begin{center}
\includegraphics[scale=0.5]{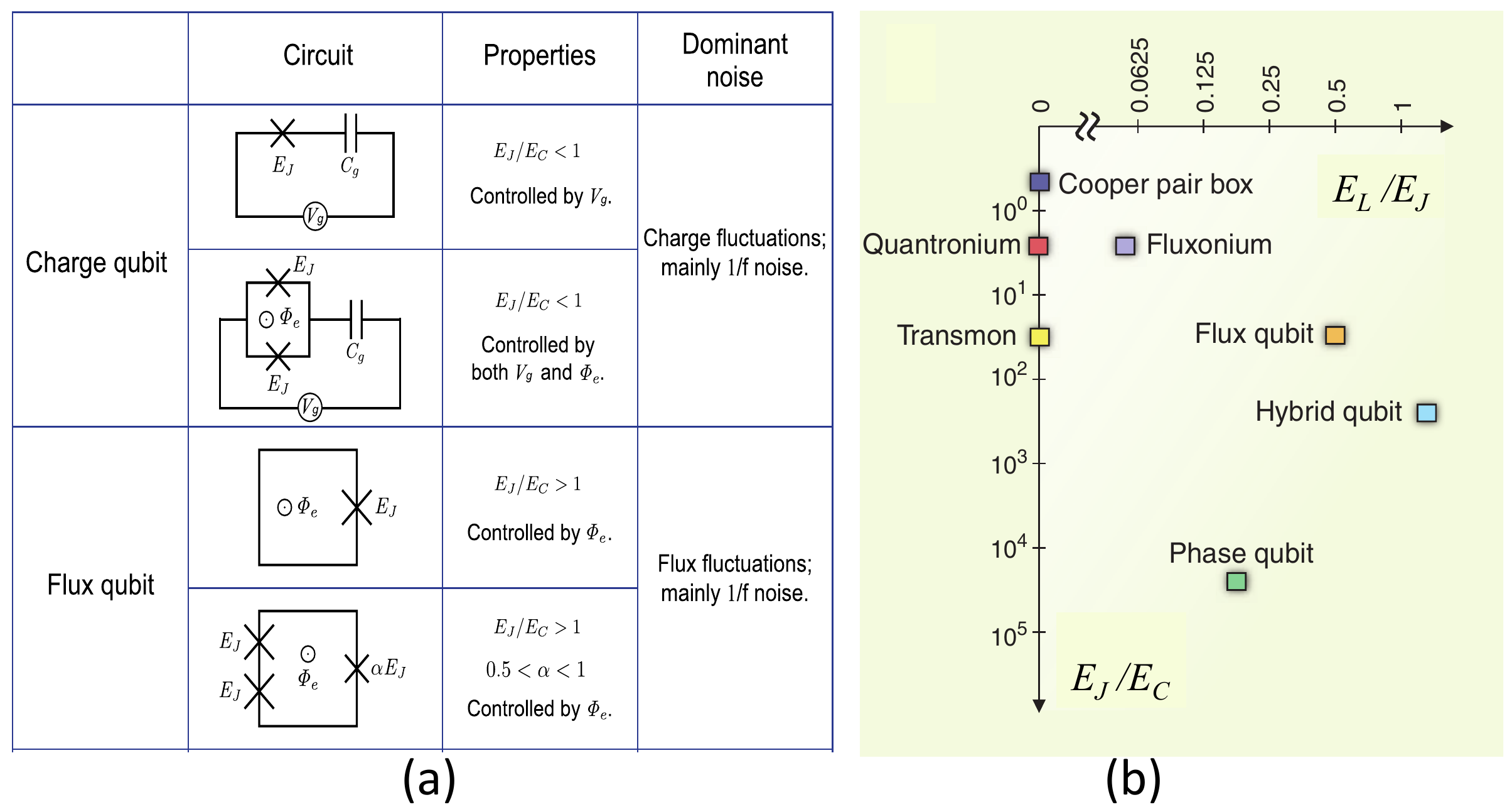}
\end{center}\addvspace{-0.5 cm}
\caption{(a)~Summary of the basic types of superconducting qubits~\cite{Xiang:2013}.
(b)~Ratios of energies $E_L /E_J$ and $E_J /E_C$ for different types of qubits 
(Mendeleev-like table)~\cite{Devoret1169}.}
\label{Fig:Classification}
\end{figure}

Several types of qubits have been realized during the last  two decades. The
simplest charge qubit, comprised of a voltage source in series with a Josephson
junction ({\it the Cooper pair box\/}), had been implemented
in~\cite{Nakamura:1999jn}. Because of the large charging energy, $E_J/E_C\ll
1$, the two charge  states different by a single Cooper pair are the working
states of this qubit. Unfortunately, the Cooper pair box is highly sensitive to
the charge noise. To overcome this difficulty, another qubit called the {\it
transmon\/} was developed~\cite{Koch:2007}. The transmon is derived from the
Cooper pair box, but it operates in a different regime of
$E_J^{\text{eff}}/E_C\gg 1$.  It benefits from the fact that its charge
dispersion and noise sensitivity decreases exponentially  with $E_J/E_C$.
Tuning $E_J^{\text{eff}}$ controls the amplitude of the potential, which forms
a periodic array of minima and maxima  shown as red and blue regions of a
contour plot in Fig.~\ref{Fig:Qubits}(b).  Since $E_J/E_C\gg 1$, tunneling
between different minima is greatly suppressed and the  qubit is realized at an
arbitrary minimum  where the lower states are unevenly spaced due to the
nonparabolicity of the cosine potential. Therefore, one can manipulate the
lowest pair of levels as in the case of a flux qubit.  In
Fig.~\ref{Fig:Classification}, we present basic types of
qubits~\cite{Xiang:2013} and show typical ratios $E_L /E_J$ and  $E_J /E_C$ for
these qubits (``Mendeleev-like table")~\cite{Devoret1169}. Selection criteria
among various qubits  for particular applications are determined not only by
internal device parameters but also by their coupling properties and tolerance
to the environmental noise.

\subsection{Inter-qubit coupling}

\begin{figure}
\begin{center}
\includegraphics[scale=0.4]{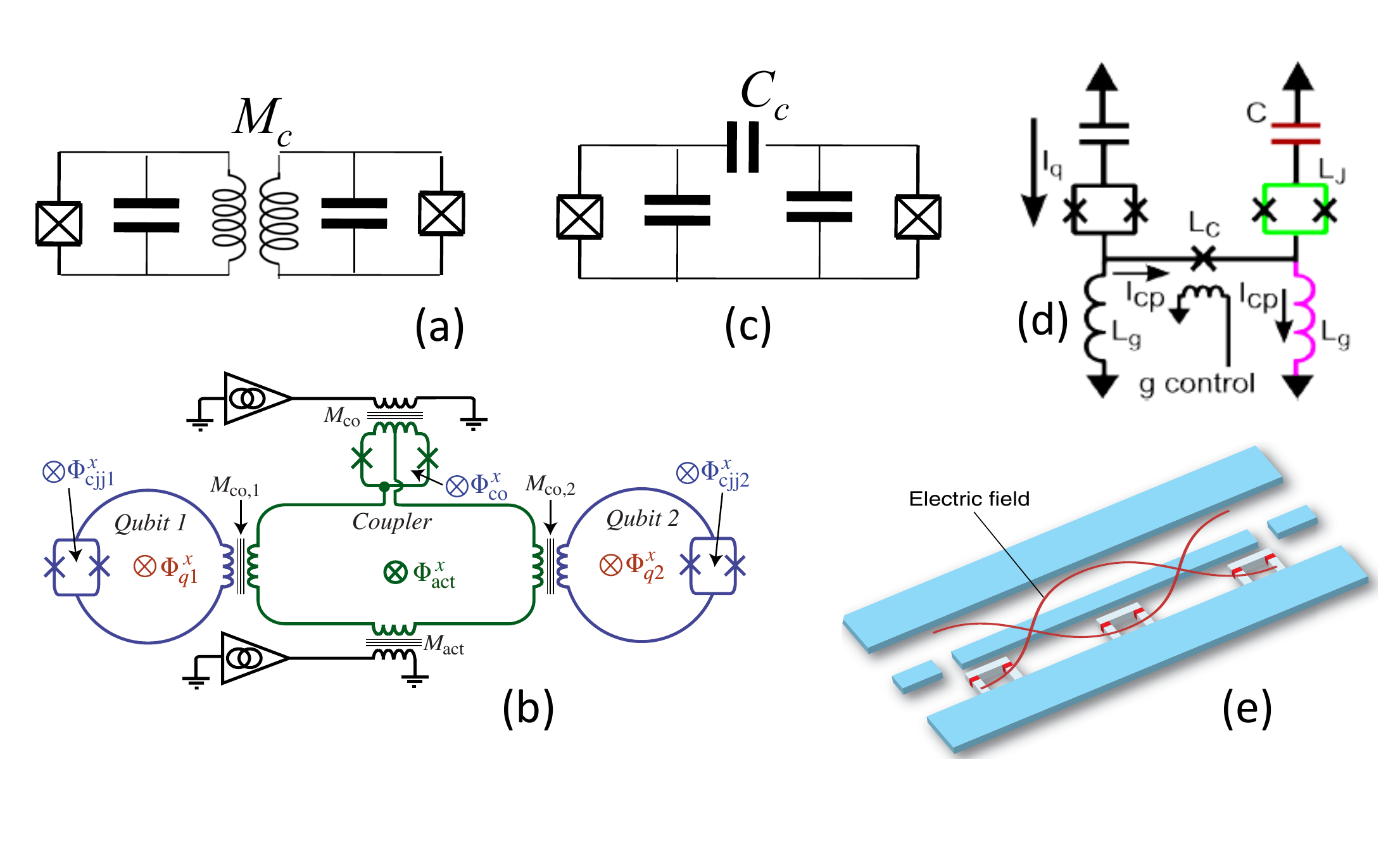}
\end{center}\addvspace{-0.5 cm}
\caption{Effective circuits for different regimes of interqubit coupling: (a)
between flux qubits via mutual inductance $M_c\equiv M_{12}$, (b) through
inductive loop controlled by SQUID, \cite{Harris:2009} (c) between transmons
coupled via capacitance $C_c$, and (d) tunable coupling between transmons
controlled by Josephson junction with nonlinear inductance $L_c$.
\cite{YuChen:2014} (e) Schematic of the coplanar waveguide resonator (light
blue), the transmon qubits and the first harmonic of the standing wave electric
field shown in red. \cite{Fink:2009,Xiang:2013} } \label{Fig:Couplers}
\end{figure}

Controllable couplings between qubits is at the heart of any quantum computing
application. The simplest and most commonly used couplers are based on linear
superconducting circuits; e.g., mutual inductances or capacitances, as shown in
Figs.~\ref{Fig:Couplers}(a) or \ref{Fig:Couplers}(b). A typical multi-qubit
system is described by an anisotropic Heisenberg Hamiltonian:
$\hat{\cal H} = \sum_{i,\alpha}  B_{i\alpha}\hat\sigma_{i\alpha} +
\sum\limits_{i,j,\alpha(i \ne j)} J_{ij}^\alpha
\hat\sigma_{i\alpha}\hat\sigma_{j\alpha}$, where  $\hat\sigma_{i\alpha}$ are
pseudo-spin Pauli matrices in a qubit $2\times2$ Hilbert space, $B_{i\alpha}$
are the components of local fields, and $J_{ij}^\alpha$ are exchange coupling
parameters.  Mechanism of inductive coupling between flux qubits $i$ and $j$
via mutual inductance $M_{ij}=M_{ji}$ (Fig.~\ref{Fig:Couplers}(a)) is
straightforward: if $M_{ij}\neq 0$, the external flux from qubit $i$ threads
through qubit $j$ loop (or vice versa) and affects the energy levels. Thus, the
longitudinal coupling (proportional to $\hat{\sigma}_{1z}\hat{\sigma}_{2z}$)
can be expressed as $J_{ij}^z\sim M_{ij} I_i I_j$. The direct inductive
coupling is not tunable; however, a tunable coupling strength can be realized
if the inductance loop is driven by a SQUID.  An example of such coupling,
utilized in D-Wave quantum annealers, is schematically shown
Fig.~\ref{Fig:Couplers}(b)~\cite{Harris:2009}. It is based on bias currents
that produce controlled flux biases.  

A circuit diagram of two capacitively-coupled transmons  is shown in
Fig.~\ref{Fig:Couplers}(c), and can be analyzed using the lumped element method
as above. As a result, the interaction Hamiltonian for a pair of transmons can
be expressed as $\hat {q}_i \hat {q}_j C/C_c$. Calculating matrix elements of
$\hat {q}_{i,j}$ within the two-level approximation, we obtain the transverse
coupling (proportional to $\hat{\sigma}_{ix}\hat{\sigma}_{jx}$) with the
coupling parameter $J_{ij}^x\sim\sqrt{\Delta E_i\Delta E_j}(C/C_{c})$, where
$\Delta E_i$ is level splitting of $i$-th transmon. The purely capacitive
coupling is not tunable, but the coupling strength can be controlled using a
non-linear coupler with Josephson junction (a tunable inductor).  This circuit
is depicted in Fig.~\ref{Fig:Couplers}(d), where arrows indicate the flow of
current for an excitation in the left qubit~\cite{YuChen:2014}. It is important
that the coupling be tunable with nanosecond resolution, making this circuit
suitable for various applications ranging from quantum logic gates to quantum
simulations. Similar circuits are employed for readout of a  flux qubit state
in D-Wave quantum annealers, where each qubit is connected inductively with a
quantum flux parametron ($rf$-SQUID with a small inductance, a large
capacitance and a very large critical current)~\cite{Harris:2010,Berkley:2010}.
Another approach is to couple all qubits to a shared passive element (quantum
bus) such as a cavity or a coplanar waveguide resonator
(CPW)~\cite{Xiang:2013}. 

\subsection{Qubit relaxation and decoherence}

Superconducting qubits are macroscopic quantum objects whose generic quantum
properties, such as superposition of states and entanglement, inherently suffer
from detrimental effects caused by a macroscopic, noisy
environment~\cite{Paladino:2014cv}.  To describe environmental noise
phenomenologically, one should take into account random charge, flux, and
Josephson junction noise sources that modulate lumped elements of the
equivalent circuit in the qubit Hamiltonians in Eqs.~(\ref{Ham2}) or
(\ref{Ham3}). 

After  tracing over the environmental variables, the qubit dynamics is governed
by the Bloch equation with two transition rates $\Gamma_1$ and $\Gamma_2$ (or
times $T_1$ and $T_2$) describing qubit {\it relaxation\/} and {\it
decoherence\/}, respectively.  The two rates are related: $\Gamma_2 =\Gamma_1
/2+\Gamma_{d}$, where $\Gamma_{d}$ describes {\it dephasing\/} due to the low
frequency noise.  The flux qubits (e.g., D-Wave qubits) studied to date suffer
from a low-frequency flux noise due to environmental
spins~\cite{Anton:2013cb,Lanting:2014cm}. This leads to a substantial dephasing
rate $\Gamma_d$ and, in turn, to a large difference between the relaxation and
decoherence rates, $\Gamma_2\sim\Gamma_d\gg\Gamma_1$. In transmon qubits, the
flux noise is absent and the low-frequency charge noise is suppressed; i.e.,
the decoherence rate is low and $\Gamma_2$ and $\Gamma_1$ are close to each
other. 

A particular choice of a qubit depends on its suitability for a given
application. For instance, quantum annealing requires strong coupling between
the qubits. Therefore, in this case the flux qubit is a preferred choice
because a typical value of the coupling parameter for D-Wave flux qubits is
several GHz. On the other hand, the coupling between transmon qubits is much
weaker (on the order of 10 MHz). Thus, the coupling and connectivity
requirements of the quantum annealing outweigh the disadvantages caused by the
higher decoherence rate of the flux qubits. 

\section{Conclusions}

The emergence of quantum annealers in the past few years has enabled the
explorations described in this paper. The next few years promise to be yet more
exciting as more sophisticated quantum annealers become available and one sees
the advent of the first universal quantum computers able to run other quantum
heuristic algorithms. The NASA QuAIL team is excited to be at the forefront of
these developments, and looks forward to working with quantum hardware and
algorithms teams from around the world to explore quantum heuristics and
thereby broaden the areas in which quantum computation has clear applications.

\section{Acknowledgements}

The authors would like to acknowledge support from the NASA Advanced
Exploration Systems (AES) program and NASA Ames Research Center.  This work was
supported in part by the  AFRL Information Directorate under grant
F4HBKC4162G001, the Office of the Director of National Intelligence (ODNI), and
the Intelligence Advanced Research Projects Activity (IARPA), via IAA 145483.
The views and conclusions contained herein are those of the authors and should
not be interpreted as necessarily representing the official policies or
endorsements, either expressed or implied, of ODNI, IARPA, AFRL, or the U.S.
Government.  The U.S.  Government is authorized to reproduce and distribute
reprints for Governmental purpose notwithstanding any copyright annotation
thereon.

\section*{References}

\bibliography{parallelComputingArticle}

\end{document}